\documentclass[conference]{IEEEtran}
\IEEEoverridecommandlockouts
\usepackage{cite}
\usepackage{amsmath,amssymb,amsfonts}
\usepackage{algorithmic}
\usepackage{graphicx}
\usepackage{textcomp}
\usepackage{url}
\def\BibTeX{{\rm B\kern-.05em{\sc i\kern-.025em b}\kern-.08em
    T\kern-.1667em\lower.7ex\hbox{E}\kern-.125emX}}
\begin{document}

\title{Impact of RAN Virtualization on Fronthaul Latency Budget: An Experimental Evaluation\\
}

\author{
    \IEEEauthorblockN{F. Giannone$^\bullet$, H. Gupta$^\star$, K. Kondepu$^\bullet$, D. Manicone$^\bullet$, A. Franklin$^\star$, P. Castoldi$^\bullet$, L. Valcarenghi$^\bullet$
}

\IEEEauthorblockA{$^\bullet$Scuola Superiore Sant'Anna, Pisa, Italy \\$^\star$Indian Institute of Technology Hyderabad, India\\ Email:francesco.giannone@santannapisa.it}}


\maketitle

\begin{abstract}
In 3GPP the architecture of a New Radio Access Network (New RAN) has
been defined where the evolved NodeB (eNB) functions can be split
between a Distributed Unit (DU) and Central Unit (CU). Furthermore, in
the virtual RAN (VRAN) approach, such functions can be virtualized
(e.g., in simple terms, deployed in virtual machines). Based on the
split type, different performance in terms of capacity and latency are
requested to the network (i.e., fronthaul) connecting DU and CU.

This study experimentally evaluates, in the 5G segment of the Advanced
Research on NetwOrking (ARNO) testbed (ARNO-5G), the fronthaul latency
requirements specified by Standard Developing Organizations (SDO)
(3GPP in this specific case). Moreover it evaluates how much
virtualization impacts the fronthaul latency budget for the the Option
7-1 functional split.

The obtained results show that, in the considered Option 7-1 functional
split, the fronthaul latency requirements are about 250 $\mu$s but they
depend on the radio channel bandwidth and the number of the connected
UEs. Finally virtualization further decreases the latency budget.
\end{abstract}


\section{Introduction}
To address the demanding requirements in terms of expected throughput,
latency and scalability, 5G networks are expected to be massively
deployed and offer an unprecedented
capacity~\cite{eric_1},~\cite{eric_2}.  A new concept of Radio Access
Network, called New RAN, has been proposed to increase performance
with limited deployment cost. In general, in the New RAN, the evolved
NodeB (eNB) functions are split into two new network
entities~\cite{3gpp}. The base-band processing is centralized in the
so-called Central Unit (CU) and the RF processing has been left at the
edge of New RAN in the Distributed Unit (DU).

The Common Public Radio Interface (CPRI), so far used to connect
BaseBand unit (BBU)(i.e., CU) and Remote Radio Head (RRH)(i.e., DU),
has shown some limitations~\cite{cipri}. CPRI is based on carrying
time domain baseband IQ samples between RRH and BBU. Thus, CPRI needs
a high capacity fronthaul, low latency, low delay variation and fine
synchronization. Guaranteeing such requirements, if Ethernet is
chosen~\cite{5GPPP} as fronthaul transport technology, is particularly
challenging~\cite{cinta,nostro,small}.

Thus new upper layer functional splits, proposed by 3GPP in TR
38.801~\cite{3gpp}, a Next Generation Fronthaul Interface
(NGFI)~\cite{ngfi}, and the new CPRI specification for 5G called
eCPRI~\cite{eCIPRI} are under definition. Different splits, however,
demand different requirements in terms of latency and capacity as
reported in 3GPP TR 38.801~\cite{3gpp}.

Moreover, recent approaches push the CU functions into the ``cloud''
(where the CU is ``virtualized''), thereby paving the way to the
so-called virtual RAN (V-RAN)~\cite{c_ran}. However, to the best of
the authors' knowledge, no evaluation has been conducted so far of the
impact of virtualization on the fronthaul latency budget.

This paper evaluates experimentally the latency and jitter
requirements for different radio channel bandwidths and different
number of User Equipments (UEs), in both physical and virtual
environment. The experimental evaluation is performed in the 5G
segment of the Advanced Research on Networking testbed
(ARNO-5G)~\cite{ARNO}. ARNO-5G allows to emulate the behavior of a 5G
network and run performance tests to evaluate several functional split
requirements. Another foreseen feature of the ARNO-5G testbed is the
possibility of virtualizing different Radio Access Network (RAN) and
Evolved Packet Core (EPC) functions to test the virtualized RAN and
EPC limits and compare them with the deployment in physical machines.

\section{The ARNO-5G Testbed}

\begin{figure}[htb]
     \centering
     \includegraphics[width=1.0\columnwidth]{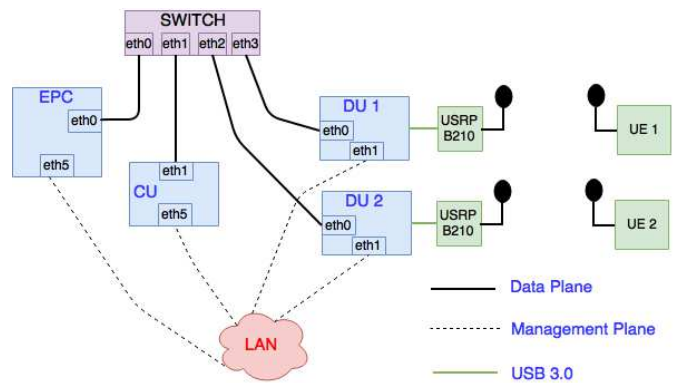}
    \caption{The ARNO 5G testbed}
    \label{fig:testbed}
\end{figure}

Fig.~\ref{fig:testbed} shows the ARNO-5G testbed. In this section the
function deployment utilized to conduct the performance evaluation
reported in this paper is described but alternative deplyoments are
possible exploiting the same hardware.

The EPC and the functional elements belonging to it (i.e., the Serving
Gateway (S-GW), the Public Data Network Gateway (PDN-GW), the Mobile
Management Entity (MME) and the Home Subscriber Server (HSS)) are
deployed in a mini-pc (Up-board) featuring an Intel Atom x5-Z8350 Quad
Core Processor and hosting Ubuntu 14.04 LTS with a 4.7 kernel
(directly precompiled by OpenAirInterface (OAI) team).

The Radio Aggregation Unit (RAU) consists of a Cisco Catalyst 2960G
switch, referred to as SWITCH in Fig.~\ref{fig:testbed}. The RAU
becomes a necessary network element because of the point-to-multipoint
architecture between the CU and the DU. The RAU forwards the
communication from the CU to several DUs.

The CU is deployed in a desktop server with Intel Xeon E5620 and
hosting Ubuntu 14.04 with 3.19 low-latency kernel. It is connected by
a 1 Gigabit Ethernet link to the EPC and by a 1 Gigabit Ethernet link
to the DU as well.

The first DU (\emph{DU1}) is deployed in a Mini-ITX
featuring an Intel I7 7700 Quad Core @ 4.0 GHz and hosting Ubuntu
14.04, 3.19 low-latency kernel. This machine is connected to the CU by
a 1 Gigabit Ethernet link. It is also connected through USB 3.0 link
to an Ettus B210 for implementing the Radio Frequency (RF) front-end.

The second DU (\emph{DU2}) is deployed on a desktop
computer with an Intel i7 4790 @ 3.60 GHz and hosting Ubuntu 14.04
with 3.19 low-latency kernel. Also the \emph{DU2} is connected through
USB 3.0 link to an Ettus B210 for implementing the RF front-end. The
Ettus B210 USRP device is a fully integrated, single-board, Universal
Software Radio Peripheral (USRP) platform and acts as radio front-end
performing Digital to Analog and Analog to Digital Conversion
(DAC/ADC).

The UEs (ie., \emph{UE1} and \emph{UE2}) consists of Huawei E3372 LTE
dongles. The dongles support LTE category 4 and Frequency-division
duplexing (FDD) communication systems in the following bands: 900 MHz,
1800 MHz, 2100 MHz and 2600 MHz.


The utilized mobile network software is the OpenAirInterface (OAI) by
Eurecom. The current OAI platform includes an implementation of 3GPP
LTE Release 10 for UE, eNB, MME, HSS, S-GW and PDN-GW on standard
Linux-based operating system. In particular, the OAI software stack of
the LTE protocol provides different layers such as PHY, RLC, MAC, PDCP
and RRC.  The latest OAI development branch was used to evaluate the
considered scenarios.

Moreover, OAI platform provides C-RAN based functional split evaluation.
The functional splits implemented by the OAI platform are
the IF5 and IF4.5 also known as Option 8 and Option 7-1 in the 3GPP
terminology~\cite{3gpp}. In our study we consider a signal bandwidth
equal to 5 MHz and 10 MHz, corresponding to 25 and 50 Physical
Resource Blocks (PRBs) with the Option 7-1 scenario. In this split in
the uplink direction, Fast Fourier Transform (FFT), Cyclic Prefix (CP)
removal and possibly Physical Random Access Channel (PRACH) filtering
functions reside in the DU and the rest of PHY functions reside in the
CU. In the downlink direction, Inverse Fast Fourier Transform (IFFT)
and CP addition functions reside in the DU, the rest of PHY functions
reside in the CU. In other word, the Option 7-1 functional split is
made before/after the resource mapping/demapping respectively.

\section{Performance Evaluation Parameters and Evaluation Scenarios}

This paper evaluates experimentally the maximum latency (i.e., the one
way delay between DU and CU) and jitter (i.e., packet delay variation)
that Option 7-1 functional split can tolerate in the fronthaul,
referred to as allowable latency budget and allowable jitter budget
respectively. For Option 7-1 split the one-way latency constraint
specified by 3GPP is 250 $\mu$s~\cite{3gpp}, mainly due to the 4 $ms$
limit of the Hybrid ARQ (HARQ)~\cite{HARQ_limit}. However, no jitter
constraint is specified.

The latency and jitter experienced along the fronthaul link is
emulated by means of the linux utility traffic control
\textit{tc}~\cite{tc_unif_dist}. The \textit{tc} utility is capable of
increasing the delay and jitter that a packet experiences on a link by
storing it in the output interface for a specified amount of time
before its transmission on the link.  A delay \textit{d0} is applied
to the ethernet interface of the machine in which the DU is deployed
and a delay \textit{d1} is applied to the ethernet interface of the
machine in which the CU is deployed. In this way a one-way latency is
inserted in the fronthaul link. For reaching the allowable latency
budget, \textit{d0} and \textit{d1} are increase with steps of 10
$\mu$s. To evaluate the allowable jitter budget instead, \textit{d0}
and \textit{d1} consists of two components: a fixed mean latency and
an additional random latency following a normal distribution whose
standard deviation is increased with steps of 5 $\mu$s. In the latter
evaluation two different scenarios are considered. In the first one we
set the mean latency close to the allowable latency budget and we
varied the jitter in order to understand if the jitter could cause a
reduction of the threshold.  In the second jitter evaluation scenario
we fixed the mean latency quite far from the allowable latency budget
and the variation of the jitter values was made to understand if
jitter could be an additional constraint for the fronthaul.  More
details about how to emulate the packet delay by using the \textit{tc}
command can be found in~\cite{tc_unif_dist}.

\begin{figure}[b]
     \centering
     \includegraphics[width=1.0\columnwidth]{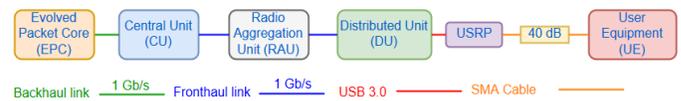}
    \caption{Scenario 1: Single DU and Single UE}
    \label{fig:testbed_single}
\end{figure}

 In the performed experimental evaluation different
scenarios are considered as described as follows.

Fig.~\ref{fig:testbed_single} shows the considered Scenario 1, where a
single DU is connected to a single CU through the RAU.  In this
scenario, we bind a single interface with a single UDP port number and
all the RAN and EPC functional elements are run on physical
machines. It is worth mentioning that NGFI can support
point-to-multipoint topology between CU and DU, thus a new element is
required. It is called RAU which can interface with CU and carries
transport for several DUs~\cite{Prototyping}.

\begin{figure}[htbp]
     \centering
     \includegraphics[width=1.0\columnwidth]{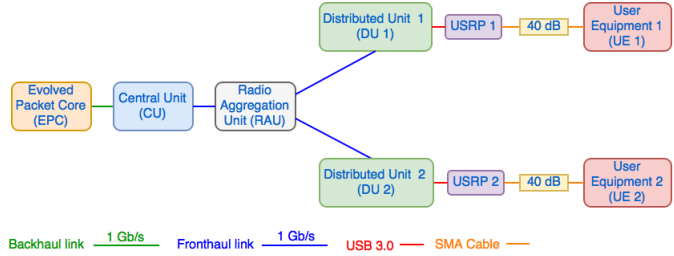}
    \caption{Scenario 2: Multiple DUs and Multiple UEs}
    \label{fig:testbed_multiple}
\end{figure}

Fig.~\ref{fig:testbed_multiple} shows Scenario 2 in which two DUs are
connected with a single CU. In order to deploy such scenario we bind a
single interface at the CU by using different UDP port numbers to
serve two different DUs at the same time. All the RAN and EPC
functional elements run in physical machines. The two DUs are
running in two different physical machines, as depicted in the block
diagram in Fig.~\ref{fig:testbed_multiple}, while two OAI CU instances,
running in the same physical machine, are connected to the corresponding
DUs.

\begin{figure}[htbp]
     \centering
     \includegraphics[width=1.0\columnwidth]{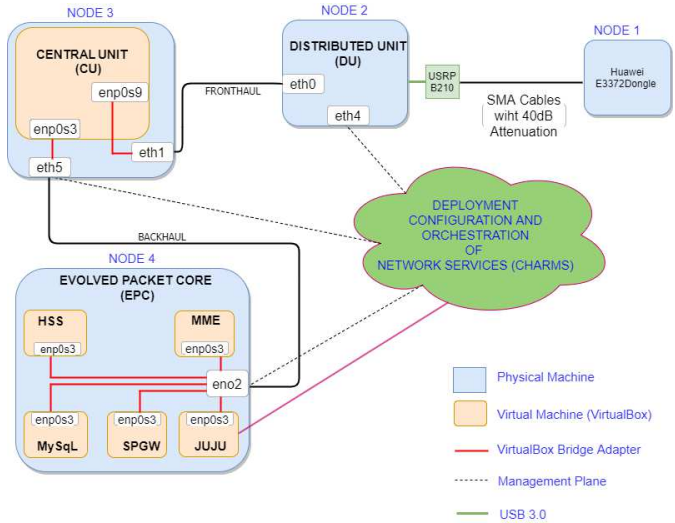}
    \caption{Scenario 3: virtualization of CU and EPC}
    \label{fig:testbed_multiple_virtual}
\end{figure}

Fig.~\ref{fig:testbed_multiple_virtual} shows the virtualized CU and
EPC setup by exploiting JuJu orchestration framework and OAI
platform~\cite{oai_juju_1}, ~\cite{oai_juju_2}. In particular, the set
of Charms (network services) managed by JuJu performs the functional
split option 7-1 as specified in 3GPP~\cite{3gpp}. This experimental
setup contains different Charms such as: MYSQL database, OAI-HSS,
OAI-MME, OAI-Serving/Packet Gateway for the EPC, OAI-eNB configured to
act as a CU and OAI-DU with attached USRP radio frequency
frontend. Each of these services is executed inside virtual machines
(running Ubuntu 16.04), exploiting VirtualBox tool, except for the OAI-DU
which runs in a physical machine, a MiniITX with Intel I7 7700 Quad Core
@ 4.0 GHz running Ubuntu 14.04.

In all the aforementioned scenarios the UEs are static and connected
to the DU through coaxial cables with 40 dB attenuation. The other
experimental parameters are shown in Table~\ref{tab:exp_pam}.
\begin{table}
\caption{Experimental Parameters}
\label{tab:exp_pam}
\begin{center}
\begin{tabular}{r@{\quad}rl}
\hline
\multicolumn{1}{l}{\rule{0pt}{12pt}
                   Parameter}&\multicolumn{2}{l}{Value}\\[2pt]
\hline\rule{0pt}{12pt}
Experiment Duration  &     100000 TTIs& \\
Frame Duration  &   10 ms& \\
Duplexing Mode  &   FDD& \\
PHY Layer Abstraction  & NO& \\
Number of DUs  & 2& \\
Number of UEs  & 2& \\
Carrier Bandwidth  & 5MHz, 10 MHZ& \\[2pt]
\hline
\end{tabular}
\end{center}
\end{table}

\section{Experimental Results}
In this section the allowable latency and jitter budgets are
evaluated. To calculate the allowable latency and jitter budgets, we
use the \emph{tc} command to add delay to network interfaces and TCP
traffic is generated by using \emph{iperf} tool to check the UE
connectivity stability.

\begin{figure}[htbp]
     \centering
     \includegraphics[width=1.0\columnwidth]{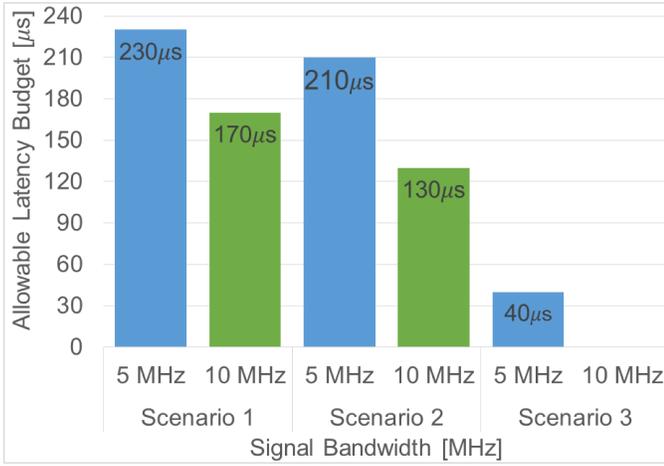}
    \caption{Fronthaul Allowable Latency Budget in the Scenario 1, Scenario 2 and Scenario 3}
    \label{fig:latency_single_multiple}
\end{figure}

Fig.~\ref{fig:latency_single_multiple} shows the allowable latency
budgets for the considered Scenario 1, Scenario 2 and Scenario 3 with
different signal bandwidth values (i.e., 5 MHz and 10 MHz). Results
show that in all the scenarios the allowable latency budget is always
below the 250 $\mu$s one-way latency constraint specified by
3GPP~\cite{3gpp}. Moreover, it can be noticed that the allowable
latency budget decreases if the signal bandwidth and if the number of
DUs connected to the same CU increases. The dependence on the signal
bandwidth is due to the heavier processing required by the higher
number of utilized PRBs. The dependence on the number of CU is
similarly due to the higher number of processes running in the same
machine. Finally, by comparing the results obtained in Scenario 3 with
the ones obtained in Scenario 1 and Scenario 2, it can be noticed that
the maximum fronthaul latency that can be tolerated is much lower if
mobile network functions are virtualized (i.e., Scenario 3) than if
mobile network functions run in physical machines (i.e., Scenario 1
and 2). This phenomenon depends on VM core capacity and other VM
parameters. Note that, in Scenario 3 with 10 MHz signal, the UE is not
capable of communicating with the EPC because the large number of
samples cannot reach the CU on time due to encapsulation delay and
transit time across the RAU.

Latency requirements for different functional splits to serve high
capacity New RAN architecture have been specified in the
3GPP~\cite{3gpp}. However, it is not clear how different functional
splits can be affected by jitter. Thus, the second set of experiments
aims at investigating whether the jitter impacts the allowable latency
budget found in the first set of experiments. In the considered
experiments, we vary the jitter while keeping the fronthaul latency
fixed and within the above allowable latency budget.

\begin{figure}[h]
     \centering
     \includegraphics[width=1.0\columnwidth]{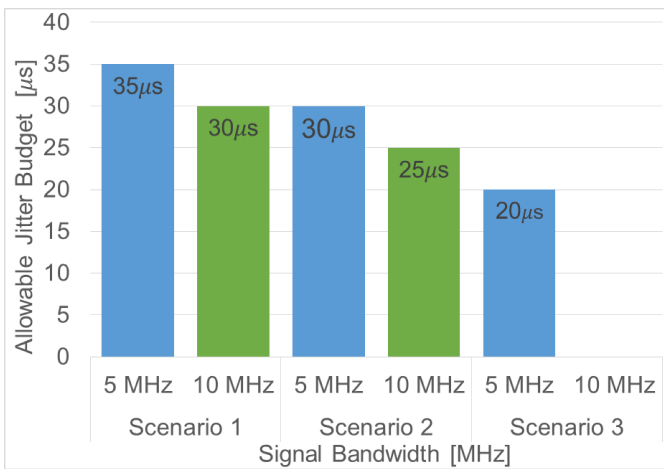}
    \caption{Fronthaul Allowable Jitter Budget with a latency close to the Limit in Scenario 1, Scenario 2 and Scenario 3}
    \label{fig:jitter_single_multiple_close_thr}
\end{figure}

Fig.~\ref{fig:jitter_single_multiple_close_thr} shows the obtained
jitter results in Scenario 1, Scenario 2 and Scenario 3. In
particular, in Scenario 1 the latency is set to 220 $\mu$s for the 5
MHz and is set to 160 $\mu$s when a 10 MHz signal bandwidth is
considered. The obtained allowable jitter budget in this case is equal
to 35 $\mu$s and 30 $\mu$s for the 5 MHz and 10 MHz signal bandwidth,
respectively.

For Scenario 2, the experiments are carried out by
setting a fixed latency on the fronthaul link equal to 200 $\mu$s and
120 $\mu$s for 5 MHz and 10 MHz, respectively. The allowable jitter
budget is equal to 30 $\mu$s for the 5 MHz and 25 $\mu$s for the 10
MHz as depicted in Fig.~\ref{fig:jitter_single_multiple_close_thr}.

In Scenario 3, the experiments are carried out by setting a fixed
latency on the fronthaul link equal to 30 $\mu$s for 5 MHz signal
bandwidth. The obtained allowable jitter budget is 20 $\mu$s. Note
that, even in this case, for 10 MHz the UE cannot connect because the
considered testbed system with virtualized CU has only 4 cores.
However, it has been verified that the UE is capable of connecting if
a virtual CU with 8 cores is utilized with no additional delay on the
fronthaul link.

By comparing the results reported in
Fig.~\ref{fig:latency_single_multiple} and in
Fig.~\ref{fig:jitter_single_multiple_close_thr} it can be deducted
that jitter negligibly impacts the allowable latency budget.

\begin{figure}[h]
     \centering
     \includegraphics[width=1.0\columnwidth]{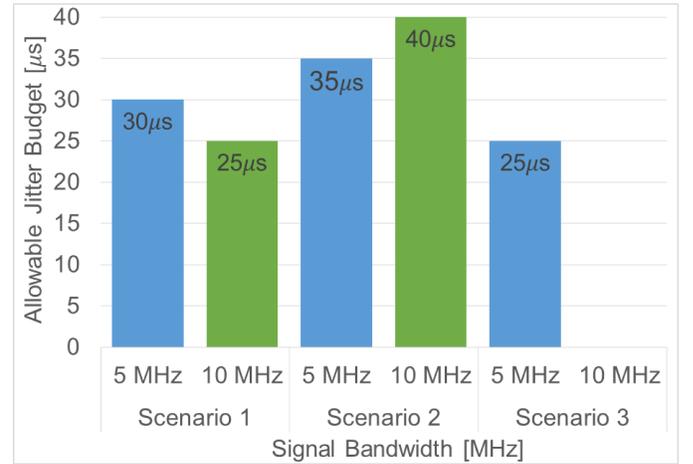}
    \caption{Fronthaul Allowable Jitter Budget with a latency from the Limit in Scenario 1, Scenario 2 and Scenario 3}
    \label{fig:jitter_single_multiple_far_thr}
\end{figure}

To observe the impact of the sole jitter on the fronthaul link, that
is to find the allowable jitter budget, the latency value is set far
from the allowable latency budget depicted in
Fig.~\ref{fig:latency_single_multiple}. The obtained results are shown
in Fig.~\ref{fig:jitter_single_multiple_far_thr} for Scenario 1,
Scenario 2, and Scenario 3. In both Scenario 1 and Scenario 2, the
latency is set to 100 $\mu$s and 50 $\mu$s for signal bandwidths 5 MHz
and 10 MHz, respectively. In Scenario 1, the obtained allowable jitter
budgets are 30 $\mu$s and 25 $\mu$s for 5 MHz and 10 MHz signal
bandwidth, respectively. Whereas, in Scenario 2, the obtained
allowable jitter budgets are 35 $\mu$s and 40 $\mu$s for 5 MHz and 10
MHz signal bandwidth, respectively. In Scenario 3, the experiments are
carried out by setting a fixed latency on the fronthaul equal to 20
$\mu$s for 5 MHz signal bandwidth. The obtained jitter budget is 25
$\mu$s, and no communication was observed in case of 10 MHz signal
bandwidth.

From the presented results, we can observe that when the jitter
overcomes a certain threshold DU and CU are not capable of
communicating. Indeed, the jitter cannot be higher than 40 $\mu s$
because, if the jitter is large, the are periods in which not enough
samples (i.e., modulation symbols) can be delivered to the PHY layer.

\section{Conclusions}
This paper presented the experimental evaluation of the impact of
virtualizing eNB functions on the fronthaul latency budget. It also
showed the maximum sustainable fronthaul jitter. The experimental
evaluation was performed in a testbed utilizing OpenAirInterface as
mobile network software, desktop computers, and USRPs.

Results showed that by increasing the instances of CU running in the
same machine the allowable fronthaul latency budget decreases of some
tens of microseconds due to the higher number of computations required
in the same machine. Similarly, but in the order of more than fifty
microseconds, it happens if the signal bandwidth increases. Moreover,
if eNB functions are run in virtual machines the allowable latency
budget further decreases, in the order of hundreds of microseconds,
due to the higher number of computations required by the
virtualization engine. Finally, the fronthaul jitter evaluation showed
that jitter negligibly impact the allowable latency budget. However,
the allowable jitter budget is in the order of tens of microseconds in
all the considered scenarios.

\section*{Acknowledgment}
This work has been partially funded by the EU H2020 5G-Transformer Project (grant no. 761536).


\begin{thebibliography}{00}
%

\bibitem {eric_1}
Ericsson, ``Ericsson mobility report,'', June 2015 (accessed on Jan. 22, 2016).
\url{https://www.ericsson.com/assets/local/mobility-report/documents/2015/ericsson-mobility-report-june-2015.pdf}

\bibitem {eric_2}
Ericsson, ``5G radio access--capabilites and technologies,'', June 2015 (accessed on Jan. 22, 2016).
 \url{http://www.ericsson.com/res/docs/whitepapers/wp-5g.pdf}

\bibitem {3gpp}
3rd Generation Partnership Project; Technical Specification Group Radio Access Network, Study on new radio access technology; radio access architecture and interfaces, 3GPP TR 38.801 V2.0.0 (2017-03).

\bibitem {cipri}
Common Public Radio Interface (CPRI) Specification V7.0, Tech. Rep., 2015 (accessed on Jun. 11, 2017).
\url{http://www.cpri.info/downloads/}


\bibitem {5GPPP}

5G PPP, ``View on 5G Architecture," White Paper, July 2016. [Online] Available: https://5g-ppp.eu/white-papers/ (accessed on Jul. 20, 2017)


\bibitem {cinta}
D. Chitimalla, K. Kondepu, L. Valcarenghi, M. Tornatore, and B. Mukherjee, ``5G Fronthaul-Latency and Jitter Studies of CPRI Over Ethernet,'' J. Opt. Commun. Netw.  9, 172-182, 2017

\bibitem {nostro}
L. Valcarenghi, F. Giannone, D. Manicone, and P. Castoldi, ``Virtualized eNB latency limits,'' in Proceedings of ICTON 2017, July 2-6, Girona, Spain

\bibitem {small}
Small Cell Forum Document 159.07.02, Small cell virtualization functional splits and use cases, Small Cell Forum, January 2016 (accessed on Jul. 20, 2017)

\bibitem {ngfi}
Next Generation Fronthaul Interface (1914) Working Group [Online] Available: \url{http://sites.ieee.org/sagroups-1914/} (accessed on Jul. 20, 2017).

\bibitem {eCIPRI}
New Common Public Radio Interface (eCPRI) for 5G Specification V1.0, Tech. Rep., 2017 (accessed on Sep. 15, 2017)
\url{http://www.cpri.info/press.html}

\bibitem {c_ran}
A. Checko, H. Christiansen, H. Yan, Y. Scolari, G. Kardaras, M. Berger, and L. Dittmann, ``Cloud RAN for mobile networks--A technology overview,'' IEEE Commun. Surv. Tutorials, vol. 17, pp. 405-426, 2015.

\bibitem{ARNO}
ARNO-5G Testbed
\url{arnotestbed.santannapisa.it}

\bibitem {HARQ_limit} CMCC, ``Transport requirement for CU\&DU
  functional splits options``, R3-161813 (document for discussion),
  3GPP TSG RAN WG3 Meeting \#93, Goteborg, Sweden, 22nd-26th August
  2016

\bibitem{tc_unif_dist}
S. Hemminger, ``Network Emulation with NetEm,'' in Linux Conf Au, April 2005.

 \bibitem{Prototyping} R. Knopp, N. Nikaein, C. Bonnet, F. Kaltenberger, A. Ksentini, R. Gupta ``Prototyping of Next Generation Fronthaul Interfaces (NGFI) using OpenAirInterface,'' [Online] Available: \url{http://www.openairinterface.org/?page_id=1695}(accessed on Sept. 28, 2017).

 \bibitem{oai_juju_1} N. Makris, T. Korakis, V. Maglogiannis, D. Naudts, N. Nikaein, G. Lyberopoulos, E. Theodoropoulou, I. Seskar, C. A. Garcia Perez, P. Merino Gomez, M. Tosic, N. Milosevic and S. Spirou ``FLEX Testbed: a platform for 4G/5G wireless networking research,'' [Online] Available: \url{https://biblio.ugent.be/publication/8506526/file/8506529.pdf}(accessed on Sept. 28, 2017).

\bibitem{oai_juju_2} ``How to run OAI using JuJu Service modelling tool (cloudification of OAI),'' [Online] Available: \url{https://gitlab.eurecom.fr/oai/openairinterface5g/wikis/OAIonJuJu}(accessed on Sept. 28, 2017).

\end{thebibliography}
\end{document}